# Space Radio Astronomy in the next $1000001_2$ years

**Leonid I. Gurvits[1]**

*Joint Institute for VLBI in Europe,*
*P.O. Box 2, 7990 AA Dwingeloo, The Netherlands*
*and*
*Department of Astrodynamics and Space Missions, Delft University of Technology,*
*Kluyverweg 1, 2629 HS Delft, The Netherlands*
*E-mail:* `lgurvits@jive.nl`

Radio astronomy and active exploration of space are peers: both began by efforts of enthusiasts in the 1930s and got a major technological boost in the 1940's-50's. Thus, for the sake of a brief review at this very special conference, it is fair to estimate the present age of these human endeavours as $1000001_2$ years*. These years saw a lot of challenging and fruitful concerted efforts by radio astronomers and space explorers. Among the high points one can mention several highly successful space-borne CMB observatories, three orbital VLBI missions, the first examples of radio observations at spectral windows hitherto closed for Earth-based observers and many yet to be implemented initiatives which are at various stages of their paths toward launch-pads of all major world space agencies. In this review I will give a "bird's-eye" picture of the past achievements of space-oriented radio astronomy and zoom into several projects and ideas that will further push the presence of radio astronomy into the space agenda of mankind over the next $1000001_2$ years. In tune with the main themes of this conference, an emphasis will be made on space frontiers of VLBI and the SKA.

---

[1] Speaker

* The author believes that the readers, devoted "digital" radio astronomers of the modern age, think binary almost by definition and recognise instantly the value of $100001_2$. But this way of thinking is not well understood by personnel departments in most European countries. They assign to the otherwise unremarkable number 65 a rather special and not necessarily adequate meaning.

# 1. Introduction

Predictions in science are known to be futile if not completely misleading. But extrapolation of the major parameters of experimental facilities and instrumentation is different and, if done correctly, can lead to efficient "strategic" decisions, especially under budgetary pressure. Just as many other modern natural sciences, radio astronomy feels this pressure quite strongly. Among all branches of modern radio astronomy, its space-oriented segment, for obvious reasons, is one of the most expensive and therefore requiring utmost strategic prudence. In this sense, good predictions literally help save resources. As shown by Freiman (1983), the dependence of the final programme cost on its preliminary estimate has a well-defined minimum, and this minimum corresponds to the most realistic estimate. In other words, in the end, you pay less if your estimate (or prediction) is as exact as possible. Yes, sometimes it is beneficial to be wise…

With this in mind, let us recall the views of space radio astronomy from its early days and try to repeat the exercise for the next $100001_2$ years.

# 2. Space radio astronomy: view from the past

Radio astronomy and active exploration of space are the brainchildren of the twentieth century. Their first steps coincided in time almost exactly and in fact were intertwined in a very peculiar way. For example, as is well known, one of the very first observations conducted by the venerable Lovell 76-m radio telescope at Jodrell Bank was aimed at detection of the second stage of the Soviet space launcher R-7 that had placed the first human-made satellite, Sputnik, into orbit in October 1957 (Lovell 1990).  Closer to the high resolution theme of this conference, the original manuscript of the first paper with the suggestion of a long baseline interferometer by Matveenko et al. (1965), according to its co-authors, contained a paragraph on the highly attractive prospects of placing one of the interferometer antennas on board a spacecraft for achieving baselines longer than the Earth diameter (N.S. Kardashev, 1985, and L.I. Matveenko, 2012, private communications). The paragraph was removed from the final version of the paper at the request of the state censorship agency: no space-related topics were allowed in non-authorised press.

Another interesting example links the early day space exploration activities and the discovery of variability of extragalactic radio sources by Sholomitsky (1965). This discovery is often seen as one of the triggers for the practical development of VLBI systems. The observations, not properly described in the original publication (again, due to the official paranoia for secrecy), had been conducted at a wavelength of 32.5 cm with the phased array of 8 parabolic antennas called ADU-1000 at the Centre of Deep Space Communications near Evpatoria (Crimea, USSR). The ADU-1000 proved to be a rather sensitive radio telescope albeit designed for very different applications – to support the "race to the Moon".

Space-based radio astronomy has many flavours. It comes with space-borne CMB observatories (Relikt, COBE, WMAP, Planck), sub-millimetre telescopes (IRAS, Spitzer, Herschel, and many others), and Space-borne VLBI (SVLBI) telescopes (TDRSS,

VSOP/HALCA and RadioAstron). The last unexplored window in the electro-magnetic cosmic emission at frequencies below 20 MHz (wavelengths longer than 15 m) is soon to be opened too. All these flavours have one characteristic in common: at least one important (usually the most important) parameter of the observing system can be realised only in space – Earth-based radio astronomy facilities cannot achieve that parameter even in principle. In the case of CMB, other sub-millimetre telescopes and Ultra-Long-Wavelength Astronomy (ULWA) this parameter is the operational wavelength (frequency). For Space VLBI systems, it is the interferometer baseline length. However evident is this reason for radio astronomy to go into space, it is a typically erroneous and sly argument by some astronomers from non-radio electro-magnetic domains too say that radio astronomy can all be done from the Earth, thus all "space" tickets should be given to high-energy astrophysicists.

A very interesting prediction of the development of space-based radio astronomy in the period 1990 – 2020 is presented in the conference proceedings "Radio Astronomy from Space" (Weiler 1987). Being currently somewhere in the second half of that period, we can compare those predictions with hard facts. This comparison is very instructive: of the 17 projects envisaged in 1987, only one (!) was launched on schedule (the Nobel–Prize fetching COBE), two were launched with 10+ and 20+ years delays (Herschel/FIRST and RadioAstron, respectively). The remaining 14 projects are either abandoned or still in the queue, 25 years after the publication of the proceedings. Interestingly, the prediction table does not contain three highly successful space radio astronomy projects implemented during the 30-year period – WMAP, VSOP/HALCA and Planck. I am not criticising the authors of the prediction table for being inaccurate: in most cases, it was not their own calculated forecast; they used data supplied by the projects themselves. But regardless of the reason for the obvious inaccuracy of the prediction, the forecast proved to be either overly optimistic or simply wrong (which is practically the same).

Having this rather discouraging example in front of us, are we brave enough to look into the next decades of the space-based radio astronomy? Let us try. Due to the conference focus on high–resolution radio astronomy and limited volume of the proceedings, attention will be given to Space VLBI, leaving a broader analysis of other radio astronomy disciplines for a discussion elsewhere.

## 3. Space VLBI in the next decades

VLBI has kept radio astronomers at the forefront of high resolution studies of the Universe for almost half a century. Over this period, several dozens of Space VLBI initiatives and proposals have been under development (Schilizzi, these proceedings). So far, only three projects have become operational. These three, plus several advanced design studies allow us to define as many as three generations of Space VLBI systems (Fig. 1). The practical history of Space VLBI (the zeroth generation) began from a brilliant ad-hoc use of the NASA communication satellite TDRSS in 1986 (Levy et al. 1986, 1989). This, by the way, is yet another example of deep intrinsic consistency between space and radio astronomy technologies: the TDRSS was designed with no radio astronomy applications whatsoever in mind. Then came the first dedicated Space VLBI mission VSOP/HALCA (Hirabayashi et al. 1998, and Hirabayashi, these proceedings), marking the first generation of SVLBI. The second and quite likely the last mission of the first generation is the RadioAstron, which enters scientific operations at the time of this conference (Kardashev et al. 2012, Kovalev, these proceedings). The two first–generation missions, VSOP/HALCA and RadioAstron, featured radio telescopes of a similar size, 8 and 10 m, respectively, and were designed to operate at frequency bands 22, 5 and 1.6 GHz (plus 327 MHz for RadioAstron). HALCA was equipped with single circular polarisation receivers, while RadioAstron can receive both senses of circular polarisation. For both missions, the data downlink operated at the rate of 128 Mbps – the limit, dictated by the technological feasibility at the end of the twentieth century. Over the zeroth and first SVLBI generations, the maximum interferometric baseline progressed from 2.2 (TDRSS-OVLBI) to ~3 (VSOP/HALCA) to ~30 (RadioAstron) Earth diameters.

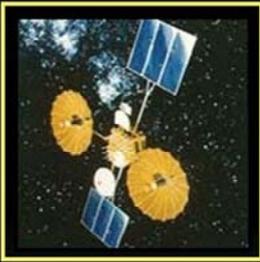

Fig. 1. Three generations of Space VLBI.

    Future developments of SVLBI will be driven by two major observing parameters, the sensitivity and angular resolution. And just as in many cases of complex engineering systems, these two parameters will be subject to complex trade-offs: one parameter can be improved only at the expense of the other one. Moreover, in addition to the engineering dimension of the trade-off, there is an astrophysical one. Namely, it might turn out to be impractical to push the resolution characteristics to the level at which the insufficient sensitivity of the SVLBI system would prevent detection of ultra-compact radio sources. So, the question is what is more important for the next generation SVLBI, the angular resolution or the sensitivity?

    By the time of this writing, RadioAstron had reported detections of continuum fringes from an AGN at 6 cm with a projected baseline of 7.2 Earth diameters and from a pulsar at 92

cm with a projected baseline of 18 Earth diameters (Kardashev et al. 2012). Hopefully, in the near future, RadioAstron will provide solid evidence for the astrophysical significance of radio structures that require baselines of the order of 30 Earth diameters and longer.

A sceptical view of the necessity to go for physically long baselines is based on the simple fact that shorter wavelengths, even with Earth-based VLBI systems, provide the same high angular resolution as SVLBI at longer wavelengths. While this argument is correct formally, its astrophysical significance is highly questionable. First of all, as known from basic physical arguments, the sensitivity of an intereferometer to the brightness temperature is linearly proportional to the physical length of the baseline, not the one denominated in wavelengths (see, e.g. Kovalev et al. 2005). Secondly, many types of compact radio sources demonstrate non-uniform spectral properties across their structure. The most typical example of this non-uniformity is the difference between flat spectra of "cores" and steep spectra of "jets" in archetypal quasars. Thus, two VLBI systems matched in angular resolution, one Earth-based operating at a shorter wavelength, and an SVLBI observing at a longer wavelength, will see astrophysically meaningful differences in the source structure at different wavelengths and maybe even tackle different mechanisms of emission (Gurvits 2000a). These two reasons justify the push for longer baselines.

The case for a higher sensitivity in the next generation SVLBI systems is obvious. The overall VLBI sensitivity depends on three main parameters: the system temperature of telescopes, the size of the interferometer antennas and the recorded signal bandwidth (the latter for continuum observations only). The first parameter for both the first generation SVLBI missions, VSOP/HALCA and RadioAstron, was already close to that of Earth-based VLBI systems. Future progress in lowering the system temperature of prospective SVLBI telescopes is feasible but limited in its impact on the overall sensitivity. The last parameter, the operational bandwidth of SVLBI systems, is defined by the maximum data rate of the down-link channel (or, more generally, the communication channel between the telescope and correlator). For both VSOP/HALCA and RadioAstron this data-rate was 128 Mbps. This value is considerably lower than the "routine" data rate of modern Earth-based VLBI systems, 1 Gbps. It is reasonable to expect that by the time the next generation SVLBI system is operational, the Earth-based VLBI systems will handle 10 Gbps per station and more. To match this value for a Space VLBI system is a big challenge. However, the good news for radio astronomers is that the technology developments needed for such the high data rate are likely to be driven (and paid for!) by industry, allowing radio astronomers to have an almost free ride.

The remaining parameter, the size of a space-borne VLBI telescope, is the subject of main concern for SVLBI planners. The fact that both first generation SVLBI missions had similar 10-m class antennas was not coincidental: it was the feasibility limit dictated by the technological and funding limitations. Several SVLBI projects with larger space-borne telescopes investigated in 1980s-1990s were not accepted primarily due to prohibitively large and expensive main antennas (Quasat, 15 m, Schilizzi et al. 1984; IVS, 25 m, Pilbratt 1991; ARISE, 25 m, Ulvestad 2000).

As of 2012, the technological and budgetary limits on the space-borne antennas size are as immutable as ever before: a deployable space-borne antenna matching in size a 25-m VLBA antenna remains a prohibitively expensive piece of hardware. Under these circumstances, it might be useful to recall an earlier suggestion to assemble a large aperture in space instead of launching it in one go (e.g. Gurvits 2000b). This way, the launch of pieces of a large space-borne structure would not be a serious challenge. Moreover, assembling large science-oriented structures might return credibility to the otherwise rather wasteful International Space Station (ISS, Fig. 2). In such the scenario, a radio telescope with a large (30 m and more) antenna can be assembled and tested while docked to an inhabitable station, and then gently accelerated toward an operational orbit using low-thrust engines. The latter is an important means for reducing the inertial load on the large antenna structure. The scenario is also attractive from the reliability standpoint: the radio telescope will be sent to its final operational orbit only after careful testing and, if necessary, fixing any problems discovered. The story of the Hubble Space Telescope that has been seriously overhauled while in orbit demonstrates the efficiency of the approach. I note that the idea of assembling a large radio telescope in orbit has been proposed more than 30 years ago (Buyakas et al. 1979).

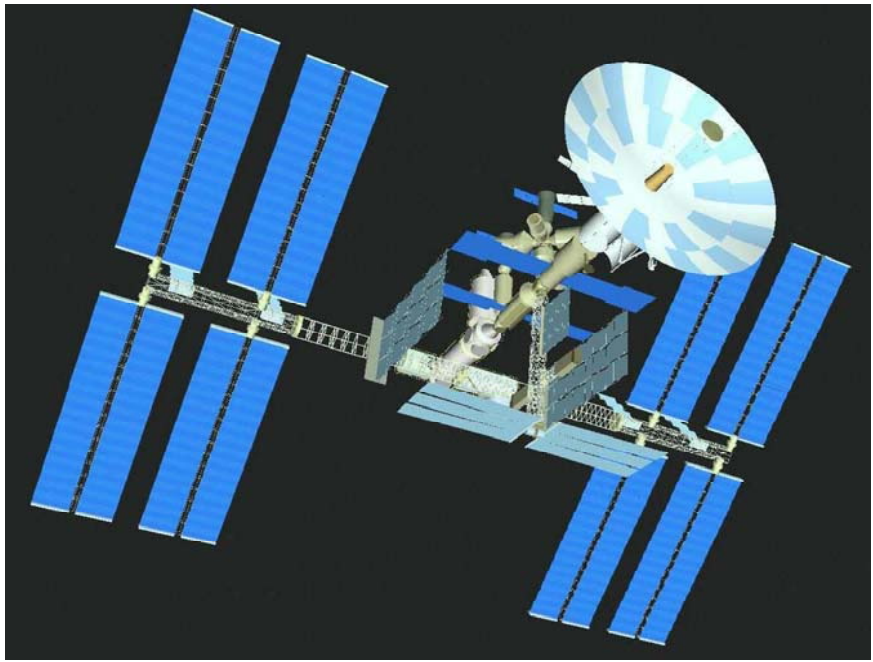

Fig. 2. An artist's impression of a large parabolic radio telescope assembled while docked to the space station. After completion of the assembly and end-to-end in-orbit VLBI tests, the radio telescope spacecraft de-docks from the station and accelerates to the operational orbit by low-thrust engines. Courtesy Aerospatiale, 1998.

At the time when the next generation Space VLBI mission will be operational, the landscape of the Earth-based radio astronomy will be dominated by large-scale facilities, such as ALMA, LOFAR/ILT and SKA. While all potential benefits of synergistic use of Space VLBI and these facilities should be addressed with great attention, the immediate impact on the choice of operational frequencies of SVLBI is obvious.

With the phased ALMA as a prospective Earth-based element of SVLBI, the space-borne telescope should be able to operate at frequencies no lower than 22 GHz. I note that coupling the phased ALMA with a space-borne radio telescope might prove to be an extremely valuable extension of the ALMA-based Event Horizon Telescope currently under development (EHT, Doeleman et al. 2009). With increasing operational frequency, the complexity of constructing a large space-borne antenna grows rapidly and is likely to limit the antenna diameter to rather modest values. For such a high-frequency SVLBI system, a very high down-link data rate might become a necessity, requiring a high-gain (and therefore large) on-board communication antenna – another difficult component of the on-board instrumentation. A possible engineering solution might be found along the lines of the concept adopted for some planetary missions (e.g. Cassini) which utilise the same antennas for communication to Earth and other ("local") tasks. Under this scenario, a medium-size high-frequency antenna conducts SVLBI observations with data stored on-board the spacecraft, and then points toward an Earth-based data acquisition station for downloading the recorded data. A somewhat similar concept of SVLBI was considered for the ad-hoc use of a communication antenna of the Venus mission VOIR (Burke 1984).

A different path and science domain might be foreseen for the combination SKA-SVLBI. The frequency domain of the early phases of SKA are "simple" for SVLBI. However, with all the attractiveness of such the system, the imbalance in size of the VLBI elements will result in severe difficulties of imaging due to the weighting problem (Garrett 2000). The remedy for this problem is obvious but expensive: the space element must be large, certainly much larger than 10-m class antennas of the first generation SVLBI. A technological solution for such a hypothetical SKA-SVLBI system might come from the SKA itself: flat antenna structures of the Aperture Array design adopted as a key technology for the SKA is most consistent with the concept of an "extendable" space-borne radio telescope mentioned above (Buyakas et al. 1979).

Table 1. Near-field radii for various Earth-based facilities and hypothetical SVLBI systems (columns) for wavelengths of 3 and 30 cm. The "1 AU" SVLBI would make possible precise 3D astrometry of pulsars in nearby galaxies.

| Baseline | 100 km | 1000 km | $10^4$ km | $10^5$ km | $10^6$ km | $10^7$ km | $10^8$ km |
|---|---|---|---|---|---|---|---|
| Facility | MERLIN | EVN-we | EVN | RadioAstron | L2 | - | "1 AU" |
| $\lambda = 3$ cm | 2 AU | 200 AU | 0.1 pc | 10 pc | 1 kpc | 100 kpc | 10 Mpc |
| $\lambda = 30$ cm | $3\times10^7$ km | 20 AU | $2\times10^3$ AU | 1 pc | 100 pc | 10 kpc | 1 Mpc |

The concepts of the next generation SVLBI discussed so far are centred around "structural" applications of VLBI. These applications have been the driving force behind the original development of the VLBI technique. But later it has been realised that VLBI is very valuable for astrometry. The same might happen to SVLBI. At present, the main difficulty in using SVLBI missions like VSOP/HALCA and RadioAstron for astrometry is related to the insufficiently accurate orbit determination. If this problem is overcome, the superior angular resolution of SVLBI might shift the emphasis of the science goals of SVLBI from structural studies to astrometry. Moreover, since the interferometer near-field radius is roughly limited by the value $B^2/\lambda$ (where *B* is the baseline and $\lambda$ the wavelength), the size of the near-field area becomes considerable and might eventually enable ultra-precise 3D astrometry of VLBI-detectable objects. Table 1 represents radii of near field areas for arbitrary wavelengths of 3 and 30 cm and various real Earth-based facilities and hypothetical SVLBI missions. The futuristic "1 AU" SVLBI will place several nearby galaxies (e.g. M31, Large and Small Magellanic Clouds) in the near-field thus enabling a complete astrometric census of their compact radio constituents, e.g. pulsars.

## 4. Concluding remarks

Space VLBI is arguably the most complex technique of modern Space science. No other technique requires such a very broad set of space-borne and Earth-based assets. These include global networks of Earth-based VLBI radio telescopes, tracking and data acquisition stations, precise orbit determination, broad-band data downlink infrastructure and specialised data processing facilities. Unlike most other types of space science missions, Space VLBI requires serial connection of devices of the on-board science package for a given frequency band. This poses a serious challenge for achieving an acceptable level of reliability: the usual remedy, redundancy, is hard to apply. This complexity drives the cost of the mission upwards, both at the development and operational phases. Not surprisingly, Space VLBI projects must conduct a very tough uphill battle while competing with other proposals for space science missions.

At this point it might be interesting to look at Space VLBI from the socio-demographic perspective. As of 2012, according to the IAU census, the total number of professional radio astronomers around the world was about 1500. The author estimates that about 400 of them have been actively involved in VLBI studies at some point or have participated in at least one Earth-based VLBI experiment. Some 50% of the latter group (this is no more than a best guess) are willing to invest their time and efforts in Space VLBI. So, the world population of Space VLBI "professionals" is no larger than 200 people. Further division into national and "territorial" domains leave no more than several dozens of Space VLBI enthusiasts per national or "territorial" space agency. Quite understandably, tight-fisted national space agencies should be convinced that catering for this rather small community is worth all the euros (dollars, yen, etc.) and efforts. Of course, one could argue that the value of science cannot be measured in terms of the number people getting their salaries as nominal scientists. True. But this kind of a posteriori trivia becomes obvious after a breakthrough discovery is made, not at the stage of building an expensive experimental facility. What is clear is that to secure a big budget for an expensive scientific endeavour like the next generation SVLBI, it is essential to couple a very strong scientific case with definitive proof of technical feasibility.

**Acknowledgements**

A brief review presented here is based on many years of close interactions of the author with hundreds of people dreaming of and working on SVLBI projects. But big teams need true leaders. Richard Schilizzi is one of them. The author is privileged to learn from Richard a great deal and work with him in many fields for *X* years. In line with the title of this paper, it is tempting to express this value *X* in the binary format. But I will simply say that at the time of this writing, $X=1000_k$, where $k > 2$; the exact value of the base, $k$, can be guessed by an initiated reader. The author is also grateful to Richard Schilizzi and Bob Campbell for very useful comments on this contribution.

## References

[1] B.F. Burke 1984, in *VLBI and Compact Radio Sources. Proceedings of the IAU Symp. 110*, eds. R. Fanti, K. Kellermann, and G. Setti, p.397.

[2] V.I. Buyakas, Yu.I. Danilov, G.A. Dolgopolov et al. 1979, *Acta Astronautica* 6, 175.

[3] S.S. Doeleman, V. Fish, A.E. Broderick et al. 2009, *ApJ* 695, 59 (doi:10.1088/0004-637X/695/1/59)

[4] F. Freiman 1983, *Transactions of the 27$^{th}$ Annual Meeting the American Association of Cost Engineers*, Philadelphia, p. 26.

[5] M.A. Garrett 2000, in *Perspectives on Radio Astronomy: Science with Large Antenna Arrays*, ed. M.P. van Haarlem, ASTRON, p. 139.

[6] L.I. Gurvits 2000a, in *Perspectives on Radio Astronomy: Science with Large Antenna Arrays*, ed. M.P. van Haarlem, ASTRON, p. 183.

[7] L.I. Gurvits 2000b, *Advances in Space Research* 26, No. 4, 739.

[8] H. Hirabayashi. H. Hirosawa, H. Kobayashi et al. 1998, *Science* 281, 1825.

[9] N.S. Kardashev, V.V. Khartov, V.V. Abramov et al. 2012, *Astron. Zh.*, in press

[10] Yu.Yu. Kovalev, K.I. Kellermann, M.L. Lister et al. 2005, *Astron. J.* 130, 2473.

[11] G.S. Levy, R.P. Linfield, J.S. Ulvestad et al. 1986, *Science* 234, 187.

[12] G.S. Levy, R.P. Linfield, C.D. Edwards et al. 1989, *ApJ* 336, 1098.

[13] B. Lovell 1990, *Astronomer by chance*, Basic Books (ISBN-13: 978-0465005123).

[14] L.I. Matveenko, N.S. Kardashev, G.B. Sholomitsky 1965, *Radiofizika* VIII, No. 4, 651.

[15] G. Pilbratt, 1991, in *Radio interferometry: Theory, techniques, and applications; Proceedings of the 131st IAU Colloquium*, San Francisco, ASP, p. 102.

[16] R.T. Schilizzi, B.F. Burke, R.S. Booth et al. 1984, in *VLBI and Compact Radio Sources. Proceedings of the IAU Symp. 110*, eds. R. Fanti, K. Kellermann, and G. Setti, p.407.

[17] G.B. Sholomitsky 1965, *Inform. Bulletin on Variable Stars*, No. 83, Konkoly Observatory, Budapest.

[18] J.S. Ulvestad 2000, *Advances in Space Research* 26, No. 4, 735.

[19] K.W. Weiler (ed.) 1987, *Radio Astronomy from Space*, NRAO Workshop No. 18,